\documentclass[sigplan,10pt,screen,nonacm]{acmart}
\usepackage{amsmath,amssymb,amsfonts,amsthm}
\DeclareMathOperator*{\argmax}{arg\,max}

\usepackage{caption}
\usepackage{pgf}
\usepackage[caption=false, font=footnotesize]{subfig}
\usepackage{graphicx}
\usepackage[inline]{enumitem}[leftmargin=*]
\usepackage[ruled, noend, linesnumbered]{algorithm2e}
\usepackage[frozencache=true,cachedir=minted-cache]{minted}
\usepackage{xcolor}

\usepackage{natbib}
\usepackage{listings}

\theoremstyle{definition}
\newtheorem{definition}{Definition}

\newtheorem{theorem}{Theorem}
\newtheorem{corollary}{Corollary}
\usepackage[short]{optidef}
\usepackage[capitalize]{cleveref}

\newcommand{\ananth}[1]{\textcolor{blue}{ananth: {#1}}}

\author{Reece Neff}
\authornote{Also with Pacific Northwest National Laboratory.}
\email{rwneff@ncsu.edu}
\affiliation{%
\institution{North Carolina State University}
\city{Raleigh}
\state{NC}
\country{USA}
}
\author{Mostafa Eghbali Zarch}
\email{meghbal@ncsu.edu}
\affiliation{%
\institution{North Carolina State University}
\city{Raleigh}
\state{NC}
\country{USA}
}
\author{Marco Minutoli}
\email{marco.minutoli@pnnl.gov}
\affiliation{%
\institution{Pacific Northwest National Laboratory}
\city{Richland}
\state{WA}
\country{USA}
}
\author{Mahantesh Halappanavar}
\email{hala@pnnl.gov}
\affiliation{%
\institution{Pacific Northwest National Laboratory}
\city{Richland}
\state{WA}
\country{USA}
}
\author{Antonino Tumeo}
\email{antonino.tumeo@pnnl.gov}
\affiliation{%
\institution{Pacific Northwest National Laboratory}
\city{Richland}
\state{WA}
\country{USA}
}
\author{Ananth Kalyanaraman }
\email{ananth@wsu.edu}
\affiliation{%
\institution{Washington State University}
\city{Pullman}
\state{WA}
\country{USA}
}
\author{Michela Becchi}
\email{mbecchi@ncsu.edu}
\affiliation{%
\institution{North Carolina State University}
\city{Raleigh}
\state{NC}
\country{USA}
}

\makeatletter
\renewcommand\noindentparagraph{\@startsection{paragraph}{4}{\z@}%
{-.1\baselineskip \@plus -2\p@ \@minus -.2\p@}%
{\z@}%
{-3.5\p@}%
{\ACM@NRadjust{\@parfont\bfseries}}}

\newcommand{\crefnames}[3]{%
  \@for\next:=#1\do{%
    \expandafter\crefname\expandafter{\next}{#2}{#3}%
  }%
}

\makeatother
\crefnames{section,subsection}{\S}{\S\S}
\usepackage{xcolor}

\newcommand{\delete}[1]{\textcolor{red}{#1}}
\definecolor{darkgrn}{rgb}{0, 0.75, 0}

\usepackage{xargs}                      
\usepackage[colorinlistoftodos,prependcaption,textsize=footnotesize]{todonotes}

\begin{document}
\title{Fused Breadth-First Probabilistic Traversals on Distributed GPU Systems}

\begin{abstract}

    Probabilistic breadth-first traversals (BPTs) are used in many network science and graph machine learning applications. In this paper, we are motivated by the application of BPTs in stochastic diffusion-based graph problems such as influence maximization. These applications heavily rely on BPTs to implement a Monte-Carlo sampling step for their approximations. Given the large sampling complexity, stochasticity of the diffusion process, and the inherent irregularity in real-world graph topologies, efficiently parallelizing these BPTs remains significantly challenging.
    In this paper, we present a new algorithm to fuse massive number of  concurrently executing BPTs with random starts on the input graph. Our algorithm is designed to fuse BPTs by combining separate traversals into a unified frontier on distributed multi-GPU systems. To show the general applicability of the fused BPT technique, we have incorporated it into two state-of-the-art influence maximization parallel implementations (gIM and Ripples). 
Our experiments on up to 4K nodes of the OLCF Frontier supercomputer ($32,768$ GPUs and $196$K CPU cores) show strong scaling behavior, and that fused BPTs can improve the performance of these implementations up to 34$\times$ (for gIM) and ~360$\times$ (for Ripples).  
     
\end{abstract}

\maketitle

\section{Introduction}
\label{sec:introduction}

\begin{figure}[t]
    \centering
    \input{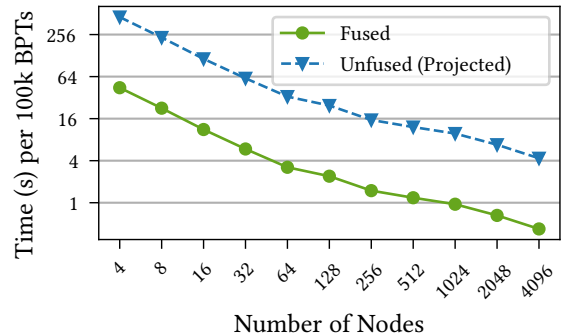}
    \vspace{-7pt}
    \caption{\small Distributed heterogeneous strong scaling on OLCF Frontier on the soc-LiveJournal1 graph for the fused implementation (this work, 64 colors) compared to an unfused baseline. 
    The chart shows: $1)$ $\sim 10\times$ speedup for fused traversals, and $2)$ strong scaling for up to $32,768$ GPUs and $196$K CPU cores.}
    \label{fig:jugaadplot}
\end{figure}

Graph traversals are a fundamental building block of graph algorithms and graph analytics~\cite{Skienna08,ExaGraph}. In particular, breadth-first searches (BFSs), probabilistic BFSs, and random walks are commonly employed in graph analytics, machine learning and deep learning~\cite{Rosvall2008,Ballal2022,Perozzi2014, Zhou2023} . For example, variants of BFS are essential to compute matchings~\cite{schrijver2003combinatorial}, network alignment~\cite{Khan2012}, and maximum flow, among many other examples. Random walks and probabilistic searches are now fundamental tools in graph representation learning, including the emerging area of graph neural networks~\cite{Perozzi2014, Zhou2023} . 

In this paper, we are motivated by the application of  probabilistic breadth-first traversals (BPT) for stochastic diffusion-based graph problems. Such applications arise in graph isomorphism tests \cite{babai1979monte} and influence maximization \cite{tang15_influen_maxim_near_linear_time}.
The stochasticity of the diffusion process is usually implemented through a Monte-Carlo sampling step that leads to performing a large number of probabilistic traversals of the graph.
For instance, in influence maximization \cite{domingos2001,kempe03_maxim}, which has numerous applications in viral marketing and computational epidemiology \cite{domingos2001,kempe03_maxim,marathe2013computational,DBLP:conf/sc/MinutoliSHTKV20}, we are interested in observing a stochastic diffusion process over an input graph in order to identify top influential nodes on the network. More formally, given a graph $G(V,E)$, a diffusion model $M$ and an integer budget $k>0$, the objective of influence maximization is to compute a seed set $S\subseteq V$ of $k$ vertices which, when activated, is likely to lead to the maximum number of activations in the graph under the diffusion model M~\cite{kempe03_maxim}. An important class of approximation algorithms for solving this NP-hard problem are based on sampling~\cite{borgs14_maxim_social_influen_nearl_optim_time, tim2014, tang15_influen_maxim_near_linear_time} (detailed in \S\ref{sec:preliminaries}). Here, each ``sample'' is a result of a single BPT on the whole input graph, and the number of samples ($\theta$) is tied to the target approximation quality. In practice $\theta$ as high as $10^5-10^6$ is necessary to to achieve an approximation close to the theoretical optimal \cite{borgs14_maxim_social_influen_nearl_optim_time,minutoli19_fast_scalab_implem_influen_maxim_algor}. 
Consequently, a dominant fraction of the runtime (up to 90\% \cite{DBLP:conf/sc/MinutoliSHTKV20}) is spent performing probabilistic breadth-first explorations of the input.


Modern supercomputers, such as Oak Ridge Leadership Computing Facility (OLCF) Frontier (currently the \#1 system on the Top 500 list and \#3 on Graph500), leverage a large number of general purpose GPUs to achieve high parallelism and high computational density. The OLCF-Frontier consists of more than 8K nodes, where each node includes 4 AMD MI250X GPUs, 
interconnected with HPE Slingshot network interfaces in a Dragonfly topology (detailed in \S\ref{ssec:Hardware}). 
Two key challenges that limit scaling BPTs on such systems are: $(i)$ the large number of BPT traversals; and $(ii)$ irregular and skewed access of memory (edges), due to probabilistic traversals and irregular structure of graph topology.

\noindent \textbf{Contributions:} 
In this paper, we introduce the technique of \emph{fused breadth-first probabilistic traversals} for distributed multi-GPU systems. This technique can be applied to any parallel use-case that executes multiple BPTs ~\cite{tang15_influen_maxim_near_linear_time,Rosvall2008,Ballal2022,Perozzi2014, Zhou2023}. 
More specifically, 
we make the following contributions. 
\begin{itemize}\itemsep=-0.05ex
\item{\em Algorithms:}
    We present the \emph{fused BPT} algorithm  that fuses various BPTs with the goal of reducing the net number of visits per edge, thereby reducing time-to-solution (\S\ref{sec:algorithm}).
    As an exemplar motivating application, we focus on  influence maximization  (\S\ref{sec:preliminaries}). 

    \item{\em Heuristics:}
    We  present several heuristics (vertex reordering, workload balancing) to improve the performance of our parallel implementation
    (\S\ref{sec:challenges}).
    \item{\em Implementations:}
    We show the efficacy of fused BPT by incorporating it into two state-of-the-art parallel influence maximization implementations, namely Ripples \cite{minutoli19_fast_scalab_implem_influen_maxim_algor} and gIM \cite{Shahrouz2021}.
    \item{\em Results:} Our experiments were conducted on 4K nodes of the OLCF Frontier supercomputer. Our results on real-world inputs show up to ~360$\times$ and 34$\times$ speedup over Ripples and gIM respectively.   We demonstrate the effectiveness of fused BPTs to decrease the number of edge accesses (\S\ref{sec:experimental-evaluation}). 
    
\end{itemize}

To the best of our knowledge, this work represents the first use of fused-BPT for influence maximization and implementation on OLCF-Frontier (Fig.~\ref{fig:jugaadplot} ), the first exascale system. We believe that this work will not only benefit the application and use of influence maximization, but also motivate the use of fused traversals in other scientific applications.  

\section{BPTs: A Motivating Application}
\label{sec:preliminaries}

As a concrete motivation for conducting a large number of parallel breadth-first probabilistic traversals, we consider the influence maximization problem. 
\begin{definition}[Inf-Max]\label{def:infmax}
Let $G=(V, E)$ be a (di)graph where $V$ is the set of vertices and $E$ is the set of edges, $M$ a diffusion process, and $k$ a budget.  The \textit{Influence Maximization Problem} is to find a set of vertices $S \subseteq V$, called seeds, such that
\begin{equation}
\argmax_{S \subseteq V} \sigma(S), \quad \text{s.t. } |S| \leq k
\end{equation}
where $\sigma(S)$ is the expected influence function over $G$ when the diffusion process $M$ starts from the seed set $S$.
\end{definition}

The problem is known to be NP-hard \cite{kempe03_maxim}. However, the expected influence function $\sigma(.)$ is a non-decreasing monotone submodular \cite{kempe03_maxim}---i.e., for subsets $A\subseteq B\subseteq V$ and a vertex $x\in V$, $\sigma(A\cup \{x\})-\sigma(A)\geq \sigma(B\cup \{x\})-\sigma(B)$. This resulted in a greedy hill climbing algorithm that provides $1-1/e$ approximation \cite{kempe03_maxim,fisher78_ii}.
An alternative class of approximation algorithms was developed using the notion of Reverse Inverse Sampling (RIS) \cite{borgs14_maxim_social_influen_nearl_optim_time}.  
The RIS algorithms use the notion of reverse reachability to assess influential vertices. In particular, RIS approaches build a collection of Random Reverse Reachable sets (RRR sets) by simulating the diffusion process $M$ in a backward manner.  The intuition is that if a vertex $u$ appears in an RRR set that was generated by starting the diffusion process at vertex $v$, then $u$ also has a chance of activating $v$ during the diffusion process; and the more number of RRR sets that $u$ appears in, the more influential it can be.  Consequently, the problem of selecting the $k$ seeds in $S$ reduces to computing a maximum-k-cover over the collection of RRR sets \cite{borgs14_maxim_social_influen_nearl_optim_time}.

The current state-of-the-art algorithm base on RIS is the IMM algorithm of \citet{tang15_influen_maxim_near_linear_time}.  \citeauthor{tang15_influen_maxim_near_linear_time} have proved a lower bound on the sample complexity (the number of RRR sets: $\theta$) that, given the size of the input graph $G$, the number of seeds $k$, and a parameter $\varepsilon$, guarantees achieving a $1-1/e-\varepsilon$ approximation bound.  
In practice, $\theta$ ranges between $10^5$-$10^6$ \cite{tang15_influen_maxim_near_linear_time,minutoli19_fast_scalab_implem_influen_maxim_algor}.
The works by \citet{Minutoli2020, minutoli19_fast_scalab_implem_influen_maxim_algor} and \citet{Shahrouz2021} provide efficient parallel implementations of this algorithm, which we use to validate our BPT fusing approach.

The network diffusion literature has generally used two simple but expressive diffusion processes: the Linear Threshold Model and the Independent Cascade Model (IC).
Under LT, the probability of a node activation depends on a group threshold parameter; whereas
under IC, the probability of a node $v$ activating its neighbors $u$ is a constant $p(e=(v,u))$ that is independent from the history of the diffusion process.
More specifically, at each step $t$, the newly activated vertices at step $t-1$ will have a single attempt at activating their 1-hop neighbors and they will succeed with probability $p(e)$. Moreover, the process assumes \textit{permutation invariant}, in that the final result of activation (or not) is independent from the relative ordering of the attempts made to activate a  vertex.  
\citet{Minutoli2020} observed that the IC model is the more computationally challenging, 
as it could lead to an irregular and often deeper propagation into the graph, among multiple concurrently advancing probabilistic BFS traversal fronts.
We now define  RRR sets under the IC model.

\begin{figure}
    \centering
    \includegraphics[width=\linewidth]{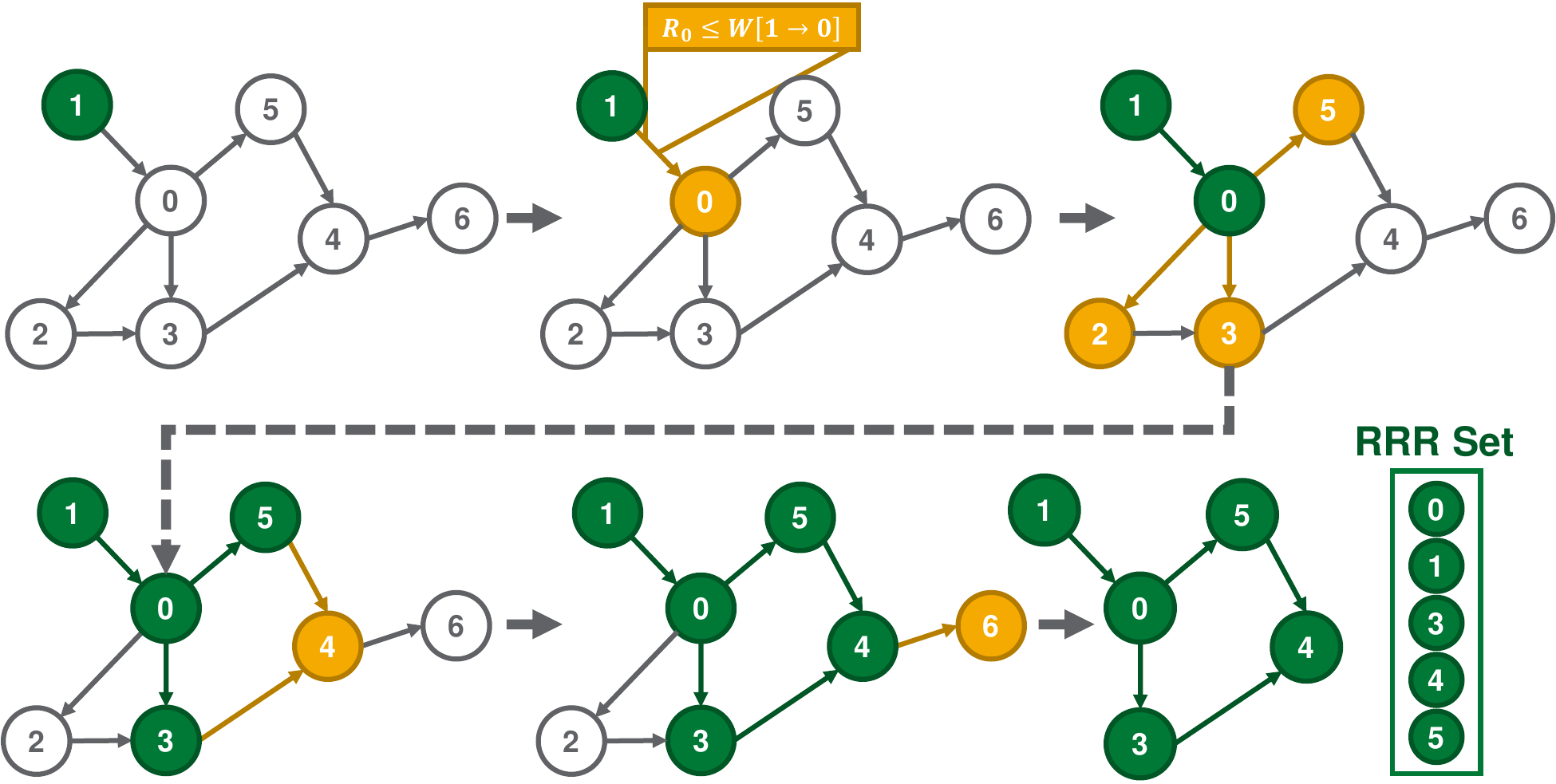}
    \caption{Example IC diffusion from a random start node.}
    \label{fig:icdiffusion}
\end{figure}
\begin{definition}[RRR set]\label{def:rrrset}
Let $v$ denote a vertex in $G$ and $\hat{G}$ denote a subgraph obtained by removing each edge $e$ of $G$ with probability $1-p(e)$. Then, the \emph{Random Reverse Reachable set} for $v$ in $\hat{G}$, denoted by $RR_{\hat{G}}(v)$, is given by:
\begin{equation}
RR_{\hat{G}}(v) = \left\{u | \textrm{ a directed path from } u \textrm{ to } v \textrm{ in } \hat{G} \right\}
\end{equation}
\end{definition}
\cref{def:rrrset} implies that RRR sets can be computed without explicitly generating the subgraphs $\hat{G}$. In fact, RRR sets can be equivalently computed as the visited array of a Probabilistic Breadth-First Traversal that visit edges with probability $p(e)$ as prescribed by the IC diffusion model. \cref{fig:icdiffusion} provides an example of the process.

\section{Fused BPT Algorithm}
\label{sec:algorithm}
To efficiently address the problem of performing a large number ($\theta$ in the case of Inf-Max) of BPTs concurrently,  we present a new algorithm, which we call the \emph{fused BPT} algorithm. 
The BPTs originate at vertices that are selected uniformly at random from $V$. 

\begin{figure}
    \centering
    \includegraphics[width=0.95\linewidth]{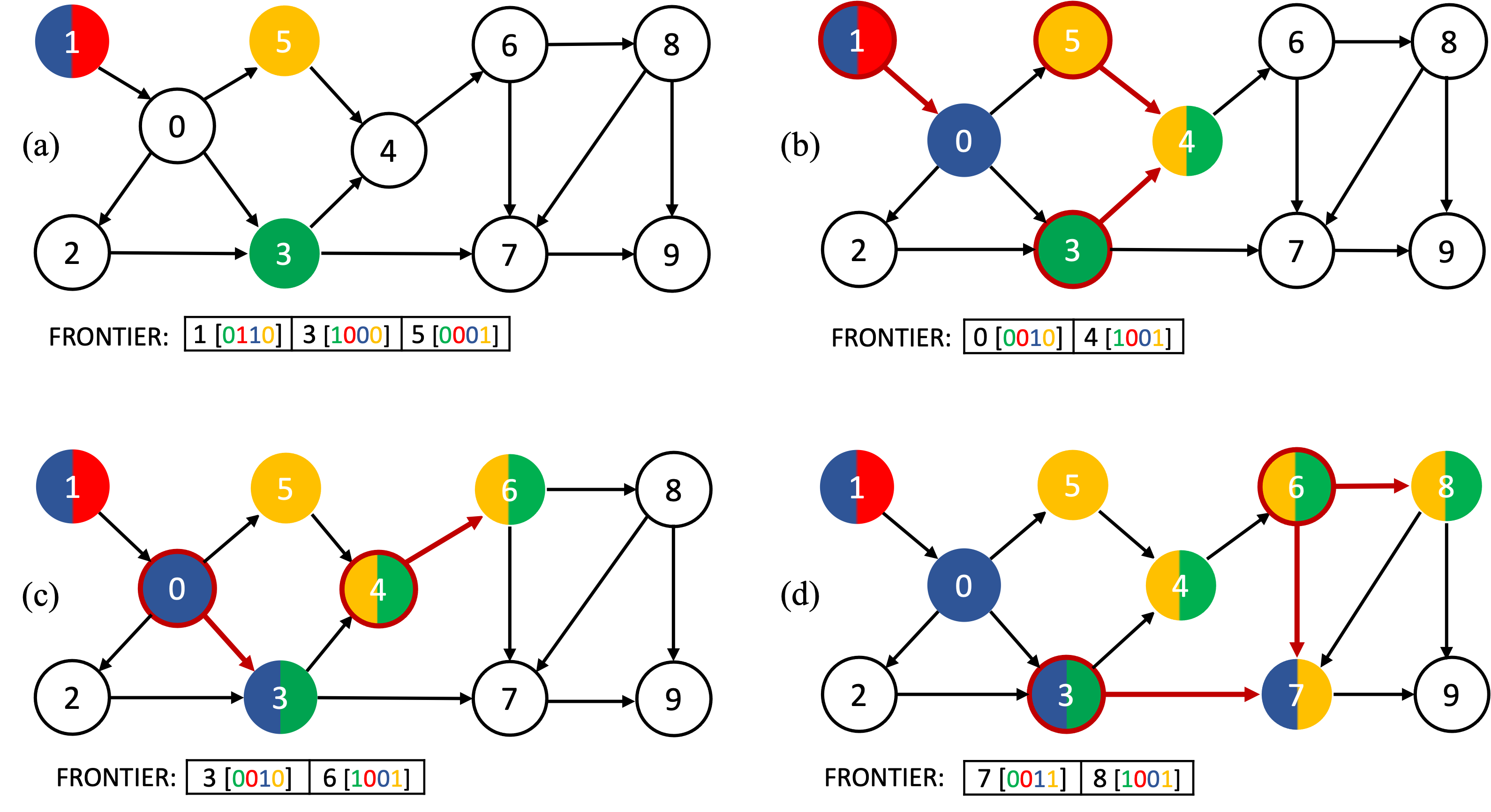}
    \caption{\small Example of fused BPT. Four BPTs originate from vertices 1, 3 and 5. The active vertices and the edges activated in each traversal step are highlighted in brown. The illustration shows the frontier \textsl{at the end} of each step.}
    \label{fig:fused-BPT-example}
\end{figure}


\noindent \textbf{Illustrative example:} Figure~\ref{fig:fused-BPT-example} illustrates the operation of the fused BPT algorithm. The example assumes four probabilistic traversals starting at vertices 1, 3 and 5. Each \emph{color} is associated to a traversal. Note that multiple traversals can originate from the same vertex (vertex 1 in the example). For each traversal step, the figure shows the frontier queue (i.e., the active vertices) at the end of that step. For each active vertex, the mask in the frontier queue shows the colors \textsl{that need to be propagated} in the next step. Due to their probabilistic nature, traversals will follow only a subset of the edges outgoing from the active vertices. Accordingly, in each traversal step only a subset of the colors is propagated. For example, in step (b) only the blue color is propagated from vertex 1 to vertex 0, while the red traversal  stops at vertex 1. 
We note that the same vertex can be traversed multiple times (as part of different traversals), leading it to be added to the frontier in different traversal steps. 
For example, vertex 3 is added to the frontier in steps (a) and (c); first time as part of the green traversal, and second time as part of the blue one. 

The key idea of fusing is as follows. When a vertex in the frontier is associated with multiple colors (as indicated by its frontier's mask), the corresponding traversals are \textsl{fused}, enabling work savings. For example, in step (b) vertex 4 is added to the frontier with two colors (yellow and green). This causes the yellow and green traversals to be fused, leading to a single traversal of vertices 6, 7 and 8. At the end of the traversal process, the vertices with the same color are associated to the same RRR set. For example, the RRR set of vertex 5 (where the yellow traversal originated) contains vertices $\{4,5,6,7, 8\}$.


\definecolor{bg}{RGB}{240,240,240}
\definecolor{codegreen}{rgb}{0,0.6,0}
\definecolor{codegray}{rgb}{0.5,0.5,0.5}
\definecolor{codepurple}{rgb}{0.58,0,0.82}
\begin{listing}
\begin{minted}[
frame=single,
fontsize=\footnotesize,
obeytabs=true,
tabsize=2,
linenos=true,
numbersep=-10pt,
highlightlines={14},
%highlightcolor=yellow
]{c++}
    /* Frontier initialization */
    for (color c)
        frontier[random(0,|V|-1)].c = 1;

    /* Fused random traversals */
    for (vertex v in frontier){
        mask fr_v = clear(frontier[v]); 
        visited[v] = visited[v] | fr_v;
        for (edge e in v.edges){
            vertex u = mate(e,v);
            mask fr_u = fr_v & !visited[u];
            for (color c in fr_u)
                if (random(0,1) > e.prob) clear(f_u, c);
            frontier[u] = frontier[u] | fr_u;  //fusing
        }
    }

    /* RRR sets construction */
    for (vertex v)
        for (color c)
            if (visited[v].c) RRRset(c).add(v);
\end{minted}
\caption{Fused BPT algorithm}
\label{listing:fused-bpt}
\end{listing}

\noindent \textbf{Pseudocode:}
Listing~\ref{listing:fused-bpt} shows the pseudocode of the fused BPT algorithm. The \texttt{frontier} array is used to identify the set of active vertices. Each element of that array is associated to a vertex and contains a bitmask that identifies the colors that need to be propagated from that vertex. If a vertex \texttt{v} is not active, \texttt{frontier[v]} does not contain any set (i.e., bit 1) colors. The \texttt{visited} array indicates, for each vertex \texttt{v}, the traversals (i.e., colors) passing through \texttt{v} (up to the current traversal step). For the example in Fig.~\ref{fig:fused-BPT-example}, the \texttt{visited} array encodes the colors of the vertices, while the \texttt{frontier} array encodes the frontier, which includes only the colors to be propagated the next time an active vertex is processed. 

During initialization (lines 1-3 of the pseudocode) a random set of vertices is selected as starting points of different traversals; accordingly, each of the selected vertices is associated a different color. The core of the traversal algorithm is encoded in lines 5-16. The traversal continues as long as the frontier is not empty. For each active vertex \texttt{v}, the corresponding frontier bitmask is read (in \texttt{fr\_v}) and then cleared (line 7). The colors in the frontier of \texttt{v} are then added to the corresponding  \texttt{visited} array to update the list of traversals passing through \texttt{v} (line 8). All the edges outgoing from \texttt{v} are then traversed with random probability (lines 9-15). For each edge \texttt{e} from vertex \texttt{v} to vertex \texttt{u}, \texttt{e.prob} indicates its traversal probability, while bitmask \texttt{fr\_u} indicates the set of colors that will be propagated from \texttt{v} to \texttt{u}. The colors already visited by \texttt{u} are excluded from the traversal (line 11), and the other colors in \texttt{fr\_v} are included with probability \texttt{e.prob} (lines 11-13). The colors in \texttt{fr\_u} are then added to the frontier of \texttt{u} (line 14), effectively fusing the newly added traversals with the ones already part of the frontier of \texttt{u}. Finally (lines 18-21) the RRR sets are updated according to the traversal outcome. 


 

\subsection{Optimality of Fused BPT} 

We analytically show that fusing BPTs does not lead to additional work, in terms of the number of edges traversed, with respect to performing unfused BPTs. 
Since sampling, fusing heuristics (sorting and reordering), and the traversals are probabilistic, we empirically demonstrate the efficacy of fusion  in \S\ref{sec:syntheticanalysis} and \S\ref{sec:experimental-evaluation}.

\begin{theorem}
\label{theorem:bpt}
The total number of traversed edges in $m$ \textit{fused} breadth-first probabilistic traversals (BPTs) is no more than the total number of traversed edges in $m$ individual BPTs.
\begin{proof}
    Without loss of generality, we will show the result for $m=2$, as well as argue for the case where the same set of edges are activated across the $m$ BPTs, regardless of whether the fusing is used or not.
    %
    We prove the lemma by induction on the depth $d$ of the traversal.  
    The base case of $d=0$ trivially holds, when both algorithms visit zero edges.
    Let us assume that the lemma is true for depth $d-1$. Specifically, the number of traversed edges from the fused BPT algorithm ($N^{(d-1)}_F$) is less than or equal to the number of traversed edges from the two separate BPTs ($N^{(d-1)}_S$) at depth $d-1$.
    
    At depth $d$, the two separate BPTs will together visit $N^{(d)}_S = N_{u_1} + N_{u_2} + 2N_{c_{12}}$ edges, where $N_{u_1}$ and $N_{u_2}$ are the edges uniquely visted by the two BPTs respectively, and $N_{c_{12}}$ is the number of edges that both BPTs will traverse. The common edges are visited only once, and therefore, the fused BPT algorithm will visit $N^{(d)}_F = N_{u_1} + N_{u_2} + N_{c_{12}}$ edges. 
    Hence, the assertion that $N^{(d)}_F \leq N^{(d)}_S$ will hold irrespective of the number of common edges traversed by the two BPTs at depth $d$, and also holds when $m > 2$.
\end{proof}
\end{theorem}

\subsection{Analysis on Synthetic Graphs}
\label{sec:syntheticanalysis}


\begin{figure*}
    \centering
    \input{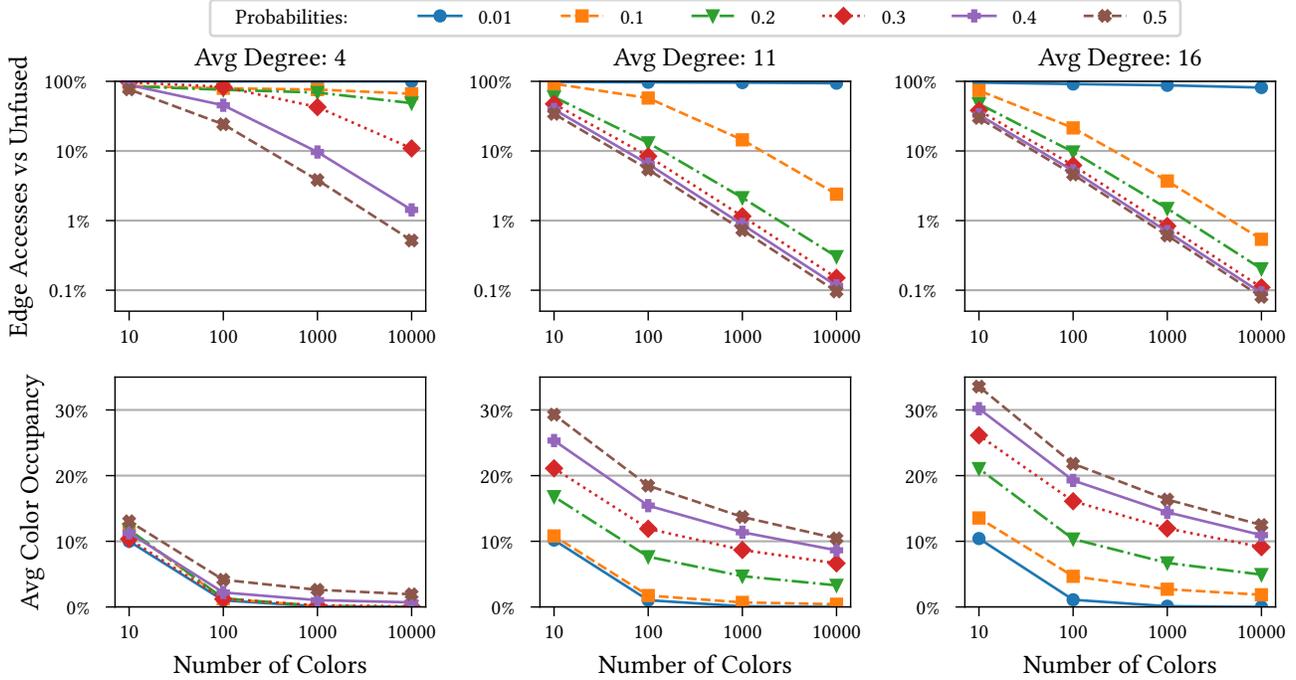}
    \vspace{-7pt}
    \caption{Edge accesses of 10,000 BPTs compared to unfused (top) and average color occupancy (bottom) for various vertex degrees, numbers of colors (i.e., group sizes), and traversal probabilities.}
    \label{fig:edge_occupancy}
\end{figure*}

Since \cref{theorem:bpt} provides a weak bound, we perform experiments to assess the amount of work saved, in terms of traversed edges, using synthetic graphs. To this end, we generate several graph configurations of the LFR benchmark, leading to graphs with vertex degrees and community sizes that follow a power law distribution \citep{lancichinetti2008}, mirroring characteristics found in real-world networks. Using NetworkX \citep{hagberg2008}, we generate graphs with 10,000 vertices and outdegrees 4, 11, and 16. For each configuration, we use three graph generation seeds. We then perform a BPT per vertex using edge probabilities 0.01, 0.1, 0.2, 0.3, 0.4, and 0.5. These traversals are repeated three times, each time using a different starting seed. This results in around 1.6 million BPTs in total.
The BPTs performed are level-synchronous (i.e., active vertices are processed level-by-level). 

We perform runs varying the number of colors from 10 to 10,000. Accordingly, BPTs are fused in groups, where the group size is equal to the number of colors used in that experiment. We then calculate the work savings (in terms of edge accesses) for each group compared to the unfused version and average across the three runs. Since we perform a level-synchronous traversal, fusing occurs only if BPTs within the same group visit a vertex in the same traversal step. \Cref{fig:edge_occupancy} shows the resulting plots. The top plot shows that higher activation probabilities and fused group sizes result in better work savings. This was expected, as higher activation probabilities result in larger activation of the graph, increasing the chances that frontiers will be shared amongst traversals. 

The bottom plots show the color occupancy, defined as the fraction of colors (i.e., traversals) that any visited vertex is part of. The average color occupancy is the average over all the vertices and traversal steps. Intuitively, this term measures how much color sharing can be exploited to fuse traversals. 
An ideal color occupancy would be as close to 100\% as possible, as this maximizes the potential for fusing.
As can be seen, although it is not feasible to fuse all BPTs within a group, the color occupancy increases with the average vertex degree and edge traversal probability. A higher color occupancy is indicative of better edge sharing within the fused group of BPTs.

\section{Fused BPT Implementation}
\label{sec:implementation}


We incorporated fusing into two existing GPU implementations of BPT extracted from gIM~\cite{Shahrouz2021} and Ripples~\cite{Minutoli2020}. Both codes perform a large number of BPTs as first step of the RIS algorithm for Influence Maximization. However, they differ in their parallelization approach. gIM performs \textit{multiple} \textsl{level-asynchronous} traversals within a single kernel. In contrast, Ripples parallelizes a \textit{single}, \textsl{level-synchronous} traversal across the whole GPU and performs multiple kernel calls for each BPT. 

\subsection{Incorporating Fused BPTs into gIM}
\label{ssec:gIM}
The baseline implementation of gIM performs multiple BPTs within a single CUDA kernel. The requested BPTs are distributed across thread blocks and performed in parallel. Each thread block consists of one warp (32 threads) and during a BPT, it processes the vertices in the frontier sequentially. For each active vertex, gIM performs edge-level parallelization and distributes the outgoing edges to the threads within the thread block. The frontier is implemented as a queue of identifiers for the active vertices, while the ``visited'' mask is an array stored in global memory---one bit per vertex. To avoid shared memory overflow,
the frontier is stored split between shared and global memory, and sections are moved between the two memory units as needed. RRR sets are implemented as a linked list of  fixed-size buffers stored in global memory. 

\vspace{1mm}
\noindent \textbf{Bugs and  fixes:} Before incorporating fusing in gIM, we fixed two existing bugs. The first bug caused gIM to lose part of the frontier queue after offloading it to global memory. The second was a concurrency bug leading some RRR sets to be generated multiple times, thus causing extra BPTs to be executed. In addition, we noticed that the large global memory utilization prevented gIM from scaling to larger graphs or inputs with larger edge traversal probabilities. To address this issue, we changed the implementation to store the RRR sets in CUDA managed memory (UVM), which enables automatic offloading onto host memory, thus limiting global memory pressure. 
When the RRR sets fit global memory, the use of UVM does not lead to performance degradation because RRR sets are written only once by the GPU kernel. 

\vspace{1mm}
\noindent \textbf{Modifications to fuse traversals:} To support fusing, we expanded the \texttt{visited} array to hold one color bitmask per vertex. For the frontier, we kept  gIM's frontier queue and added an extra global memory array to store the bitmasks associated to the active vertices (similar to the \texttt{frontier} array in \cref{listing:fused-bpt}). Fusing allows \texttt{NCOLORS} traversals to be performed in a synchronous fashion, with \texttt{NCOLORS} being the number of colors used. Accordingly, in the fused implementation each thread block performs \texttt{NCOLORS} BPTs while processing a single frontier queue. To keep the relative use of shared memory per BPT unmodified, we increased the thread block size from one to \texttt{NCOLORS} warps. With this increased block size, however, edge-level parallelization over the whole thread block can lead to thread underutilization, especially for low outdegree vertices. To maintain the same thread utilization as in the original gIM, we assigned to each warp a different active vertex, effectively parallelizing the frontier's processing. We note that edge-level parallelism can cause warp underutilization for low outdegree vertices even with this vertex-to-warp mapping. To this end, we tested the use of CUDA cooperative groups to implement finer-grained parallelization where one vertex is assigned to only 16 or 8 threads. In our experiments, however, vertex assignment at a sub-warp granularity reported a (slight) performance improvement only on the smallest graph considered. We note that the use of bitwise operations and integer intrinsic on 32 or 64 bit masks allows for efficient color processing. On the other hand, further parallelizing the for-loop at line 12 of \cref{listing:fused-bpt} by distributing colors across threads would cause thread under-utilization (when only few colors are set) and require extra synchronization on the bitmasks' updates, negatively affecting performance.
In the RRR sets construction step (lines 18-21 of \cref{listing:fused-bpt}), each thread processes an element of the \textsl{visited} array and updates the corresponding RRR sets atomically, leading to some synchronization cost. Finally, we observed that limiting the register utilization to 32 registers per thread allows increasing the GPU occupancy by doubling the number of resident thread blocks per SM, improving performance despite some added register spilling.

\subsection{Incorporating Fused BPTs into Ripples}
The baseline implementation of Ripples 
distinguishes 
each CPU core as a CPU worker or a GPU worker. A CPU worker is responsible for performing its own BPT, while a GPU worker handles kernel launches and memory movement between the host and GPU. Both kinds of workers perform a single BPT at a time. To handle the irregular workloads common in BPTs, an atomic variable denoting the number of required BPTs is located in the host, with each CPU and GPU worker performing an \texttt{atomicAdd} before performing a BPT to determine if there is more work.

Compared to gIM's threadblock granularity, Ripples utilizes the whole GPU to perform a single level-synchronous BPT, returning to the host between kernel calls. There are two main steps in the traversal:
\begin{enumerate*}
    \item \textsl{Frontier Generation:} The host launches a device-wide kernel to perform an XOR between the visited and frontier arrays to determine which BPTs are in the frontier queue, and reduces these frontier nodes to one of four bins, with each bin corresponding to a different range of degrees. The frontier array is then written to the visited array.
    \item \textsl{Edge Traversal:} The host then launches four separate streams, each with varying levels of thread block granularity, to scatter the edges and perform randomized traversal. Nodes with smaller degrees have all their edges traversed by a single thread within the warp, and nodes with larger degrees have all their edges traversed by a thread block of size 32, 256, or 1024. This implementation is similar to \citet{Merrill2012}.
\end{enumerate*}
Afterwards, the visited array is moved to the host, and the host CPU traverses the array to push the visited vertices to the appropriate vector.

\vspace{1mm}
\noindent \textbf{Modifications to fuse traversals:}
To perform fusing on the baseline implementation of Ripples, we turned the \texttt{visited} and \texttt{frontier} arrays (lines 7-8 of \cref{listing:fused-bpt}) into a blocked bitmask, with each block containing \textit{N} 32-bit values, where \textit{N} contains a block of 32 colors, one bit per color. This blocking ensures proper memory alignment between warps. Each bit within the block corresponds to a different color, or BPT, within the fused group. Increasing the number of variables, however, resulted in threads handling colors across multiple variables, which could lead to memory access inefficiencies. To alleviate this, we assign \textit{N} threads per blocked bitmask where, as in fused gIM, each thread utilizes integer intrinsics to process their color (line 12 of \cref{listing:fused-bpt}). Because the number of colors for all edges are the same, there is no workload imbalance in the warp/block-level hierarchical queue. During frontier queue generation, where vertices are filtered to a degree-based workload queue, we perform a localized warp-level reduction to determine the queue offset. Then, the leading thread broadcasts the proper offset to each thread to reduce the number of global atomic operations.

\vspace*{-0.1in}
\section{Parallel Heuristics}
\label{sec:challenges}




\noindent \textbf{Vertex Reordering Techniques:} 
\begin{figure}
    \centering
    \input{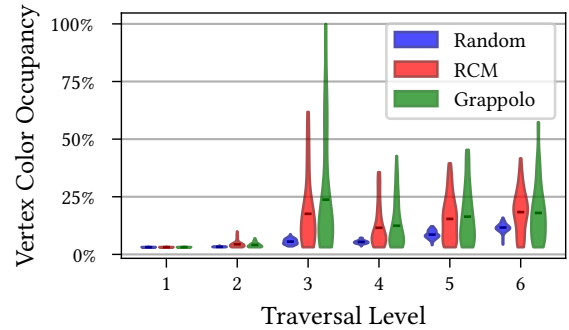}
    \vspace{-2.5mm}
    \caption{Color occupancy on a fused web-BerkStan traversal with 32 colors using RCM and Grappolo (\S\ref{reordering}).}
    \label{fig:reordering}
\end{figure}
\label{reordering}
One of the key factors that determines the parallel performance of fused BPTs is the degree to which we can visit shared vertices between fused BPTs around the same time during execution. We refer to this as the locality of fused BPTs. For example, in Figure~\ref{fig:fused-BPT-example}(b), the yellow and green BPTs from vertices 5 and 3 respectively, need to converge on the shared vertex 4 around the same time in order to benefit from fusing. However, if in memory these vertices are stored in a non-local fashion, then such probability of shared visits in time is reduced. Therefore, to help increase the probability of sharing while fusing, maintaining vertex locality is important.


A classical technique for improving vertex locality is vertex reordering, which aims to obtain a locality-preserving permutation.
Reordering algorithms aim to either explicitly minimize the gap (e.g., minimum linear arrangement algorithm) or use heuristics (e.g., reverse Cuthill-McKee (RCM), degree-based sorting and partitioning-based sorting) to accomplish similar goals in an efficient manner~\cite{Barik2020}. 

Our motivation for employing vertex reordering is to maximize locality for fused BPTs in a given batch, and in the process increase vertex color occupancy. We experimented with several heuristics from ~\citet{Barik2020} and observed that clustering-based heuristic (Grappolo) provides the best performance. As an example, comparison for web-BerkStan at different traversal levels is summarized in \cref{fig:reordering}. We observe significantly larger color occupancy for both RCM and Grappolo relative to a randomized vertex reordering as the baseline. All three schemes consider the sorted variant, which pre-generates and sorts the random start vertices. When workers retrieve BPTs to fuse, they pull from this sorted list of start vertices for better locality and, consequently, more opportunities for fusing.

\begin{figure}[!hbt]
    \centering
    \input{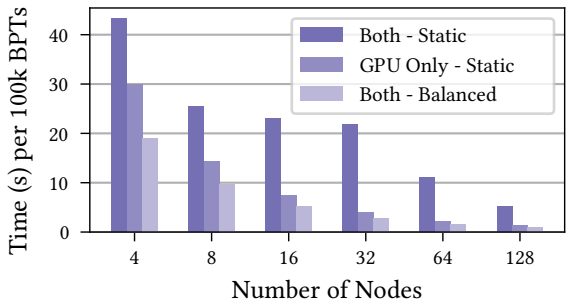}
    \vspace{-7pt}
    \caption{Multi node scaling on soc-pokec-relationships.}
    \label{fig:scalingbalancing}
\end{figure}

\noindent \textbf{CPU-GPU Workload Balancing:}
Originally, the GPU and CPU workers of Ripples would obtain a color size---meaning a single node would process up to 3,584 BPTs at a time, if the color size is 64 (56 BPT groups). When scaling to multiple nodes, the number of BPTs each node needed to generate was scaled down proportionately. However, we encountered an issue where the heterogeneous CPU-GPU setup was lacking in performance compared to a GPU-only setup (Fig.~\ref{fig:scalingbalancing}). Upon examination, we found that the CPU workers were causing workload starvation, with a comparative test run on web-BerkStan showing the CPU implementation to be up to $16\times$ slower than the GPU implementation. 

To alleviate this issue, we designed a lightweight micro benchmarking scheme where, at the beginning of each run, the host times several batches to be run by each CPU and GPU worker. Then, the host calculates the difference between the average CPU and GPU worker times. This difference is split, increasing/decreasing the CPU worker color size, until the timings between CPU and GPU workers are similar. 
This method worked for smaller graphs, but in larger graphs, the micro benchmarking was setting the CPU color size to 0, i.e., the CPU would cause starvation even when retrieving a single BPT. To enable the CPU workers to assist with BPT generation even if a single core would cause starvation, we group CPU workers in the same L3 cache region (6 CPU cores each) to collaborate on one BPT group, resulting in 8 total CPU worker groups executing BPTs.
The results of this workload balancing are shown \cref{fig:scalingbalancing} (see \cref{sec:ripplesscaling}).

\vspace*{-0.1in}
\section{Experimental Setup}
\label{sec:experimental-setup}



\begin{table}[!th]
    \caption{SNAP Graphs}\label{table:graphs}
    \begin{center}
    \resizebox{\linewidth}{!}{
        \begin{tabular}{lrrr}
        \toprule
            \textbf{Graph} & \textbf{\# Nodes} & \textbf{\# Edges} & \textbf{Avg. Degree}\\
        \midrule
            web-BerkStan (BS) & 685,230 & 7,600,595 & 22.18\\
            web-Google (Go) & 875,713 & 5,105,039 & 11.66\\
            soc-pokec-relationships (PR) & 1,632,803 & 30,622,564 & 37.51\\
            wiki-topcats (TC) & 1,791,489 & 28,511,807 & 31.83\\
            com-Orkut (Ok) & 3,072,441 & 117,185,083 & 76.28\\
            soc-LiveJournal1 (LJ) & 4,847,571 & 68,993,773 & 28.47\\
        \bottomrule
        \end{tabular}
        }
    \end{center}
\end{table}

\noindent\textbf{Hardware:}
\label{ssec:Hardware}
We study the performance and scalability of Ripples on Crusher and Frontier Computing Systems hosted at Oak Ridge Leadership Computing Facility. These systems have the same configurations, where each compute node has a 64-core AMD EPYC 7A53 CPU, 512 GB DDR4 memory, and 4 AMD MI250X GPUs. Each MI250X contains two Graphics Compute Dies (GCDs) that, to the host runtime, appear and operate as two separate GPUs. For this reason, we refer to a GPU as a single GCD. We perform scaling experiments up to 4,096 compute nodes, utilizing HPE Cray MPICH 8.1.23 for multi-node setups. At the time of this writing, Frontier is the \#1 system on the Top500 list and \#3 on the Graph500 list.

Our experiments are run with “Low-Noise mode” enabled to minimize operating system noise. This setting restricts system processes to run on the first core of each L3 cache region (8 L3 regions in total). However, using the low-noise mode implies that applications are restricted to using only the remaining 56 CPU cores on each node.
The gIM framework supports only a single NVIDIA GPU. For our experiments, we use a Tesla A100 GPU with 40 GB of HBM2 memory. 

\vspace{1mm}
\noindent\textbf{Software:}
We have implemented the fused BPT approach in gIM using CUDA 12.1 and GCC 11.3.0.
For Ripples, we implemented the fused BPT approach using AMD HIP for GPUs and OpenMP for CPUs. The host code for scheduling work on GPUs and CPUs also uses OpenMP. We compiled Ripples with hipcc (from AMD ROCM 5.1.0). \textit{A git repository will be released upon acceptance to preserve anonymity.}



Unless otherwise specified, we assigned edge weights  from a uniform distribution between 0 to 1. We performed the graph generation process once, and consistently reused the same inputs. For multi-node setups, workload balancing as discussed in \cref{sec:challenges} is performed as a pre-processing step.





\vspace*{-0.1in}

\section{Experimental Evaluation}
\label{sec:experimental-evaluation}

\begin{figure*}
    \centering
    \input{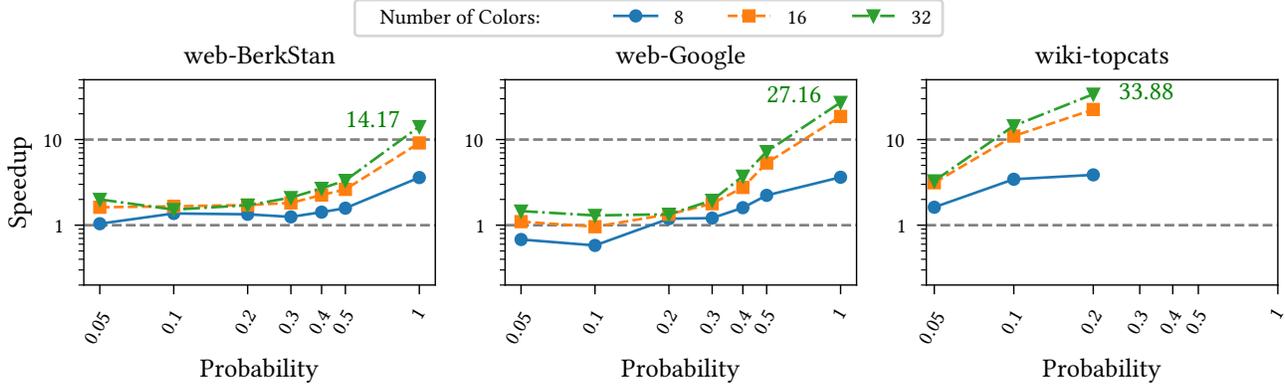} 
    \vspace{-2.5mm}
    \caption{Speedup of fused over unfused gIM with different numbers of colors and traversal probabilities. Missing data points correspond to experiments where gIM run out of memory (despite using UVM for RRR sets). 
    }
    \label{fig:gim_fused}
\end{figure*}

\subsection{gIM with Fused BPTs}
\label{sec:results:gim}

Fig.~\ref{fig:gim_fused} shows the effect of fusing BPTs on gIM's performance for three of the graphs in Table \ref{table:graphs}, namely, web-Berkstan, web-Google, and wiki-topcats. 
We do not show results for soc-pokec-relationships, soc-LiveJournal and com-Orkut because gIM  run out of memory on those graphs. We recall that we used UVM only for RRR sets, while we kept the original gIM implementation for dynamically allocated data structures (such as the frontier queue). We conducted experiments with various edge traversal probabilities and numbers of colors. 
We make the following observations.

First, incorporating fusing of BPTs in gIM is beneficial in most cases, and yields speedups up to $14\times$, $27\times$, and $33\times$ on web-Berkstan, web-Google, and wiki-topcats, respectively. The performance benefits are mainly due to two advantages of the fused gIM implementation. First, fusing decreases the number of vertices and edges traversed, also reducing the number of accesses to global memory. Second, since fused-gIM uses a single frontier queue for all BPTs in the same group, fusing can reduce the probability of overflowing the shared memory allocated to the frontier queue, thus limiting the data movements between the shared and global memory. 


Second, in the absence of enough fusing opportunities, 
the \textit{overhead of the extra code} added to incorporate fusing (see \S\ref{ssec:gIM}) can lead to performance degradation
over the baseline gIM. For example, when using only 8 colors and traversal probabilities below 0.2, web-Google incurs a  $32\%$ performance degradation. However, as the fusing opportunities increase (e.g., higher traversal probabilities), fusing BPTs can considerably improve gIM's performance. 

Third,  increasing the \textit{number of colors} from 8 to 32 results in a significant performance improvement.
We recall that the number of colors determines the size of each BPT group, i.e., each set of BPTs that can be potentially fused. Larger groups allow more color sharing opportunities, enabling more fusing.
For example, on  web-Berkstan  fusing 32 BPTs results in a speedup up to $14.7\times$, while fusing 8 BPTs results in a maximum speedup of $3.6\times$. 

Furthermore, lower \textit{traversal probabilities} reduce the chance of the same edge being traversed by multiple BPTs. This hinders the performance gains from fusing. For example, when using 32 colors on wiki-topcats, increasing the traversal probability from $0.05$ to $0.2$ causes the speedup over the unfused gIM to increase from  $3.2\times$ to $33.8\times$.

Finally, we  evaluated \textit{the impact of sorting the starting vertices} of the BPTs. While sorting helps the performance of web-Berkstan, bringing the maximum speedup over the baseline from $14.1\times$ to  $23.9\times$, it does not report additional performance gains on the other graphs. In most cases,  the overhead of sorting the starting vertices outweighs the benefits from increased fusing opportunities.


\begin{figure*}[tbh]
    \centering
    \input{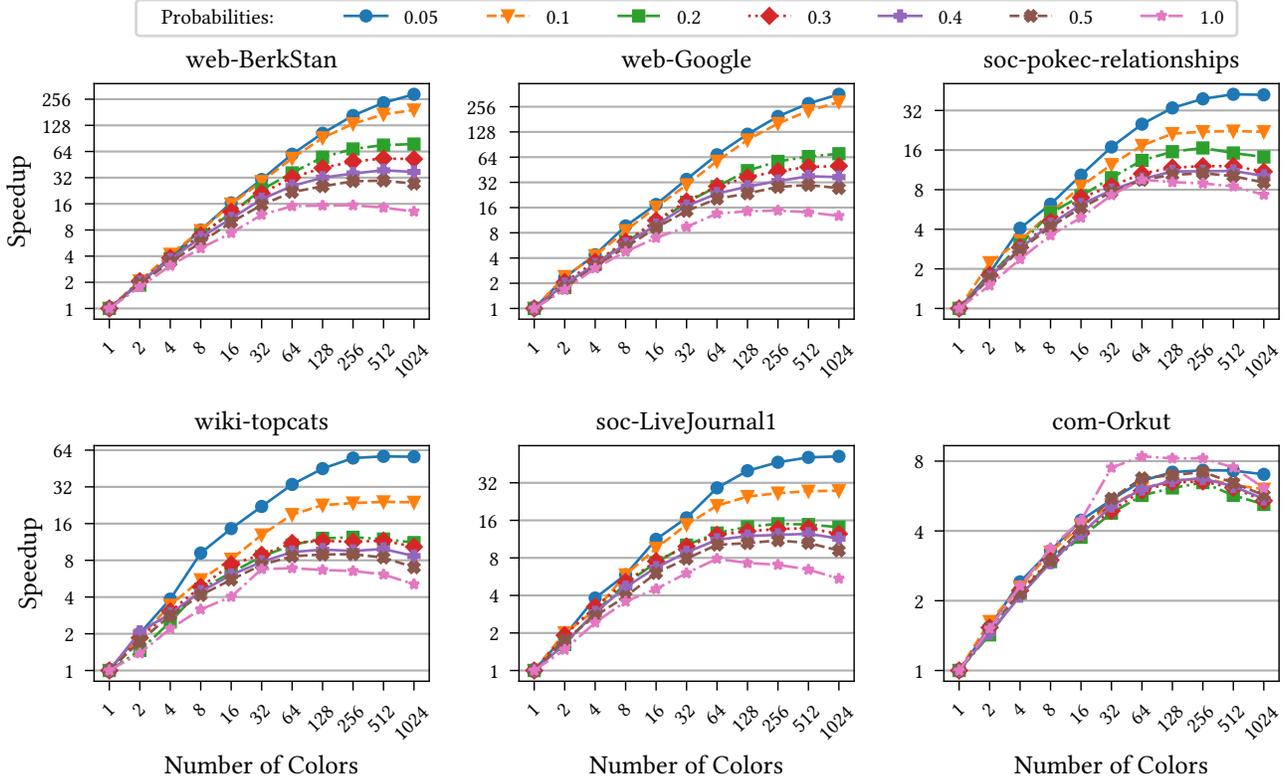}
    \caption{Speedup of fused over unfused Ripples when varying numbers of colors and traversal probabilities.}. 
    \label{fig:probscaling}
\end{figure*}

\subsection{Ripples with Fused BPTs}
\label{sec:results:ripples}
The Ripples framework approaches the problem of BPT generation from a device-wide perspective. As such, we evaluate the impact of fusing on Ripples, including its behavior on single- and multi-node scaling.

\subsubsection{Sensitivity Analysis}


\begin{figure}[b]
    \centering
    \input{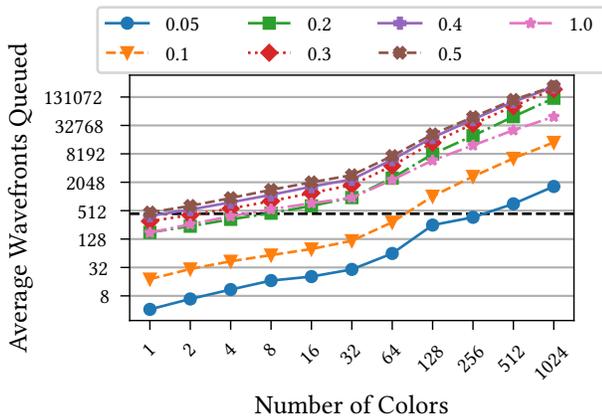}
    \caption{Frontier profiling of 10K BPTs on web-BerkStan. The y-axis denotes the number of wavefronts queued per iteration of the BPT. The horizontal black dashed line denotes the number of SIMD units present on one GCD of the MI250x (440), where each SIMD unit processes one wavefront.}
    \label{fig:probutilization}
\end{figure}

\begin{figure*}[!hbt]
    \centering
    \input{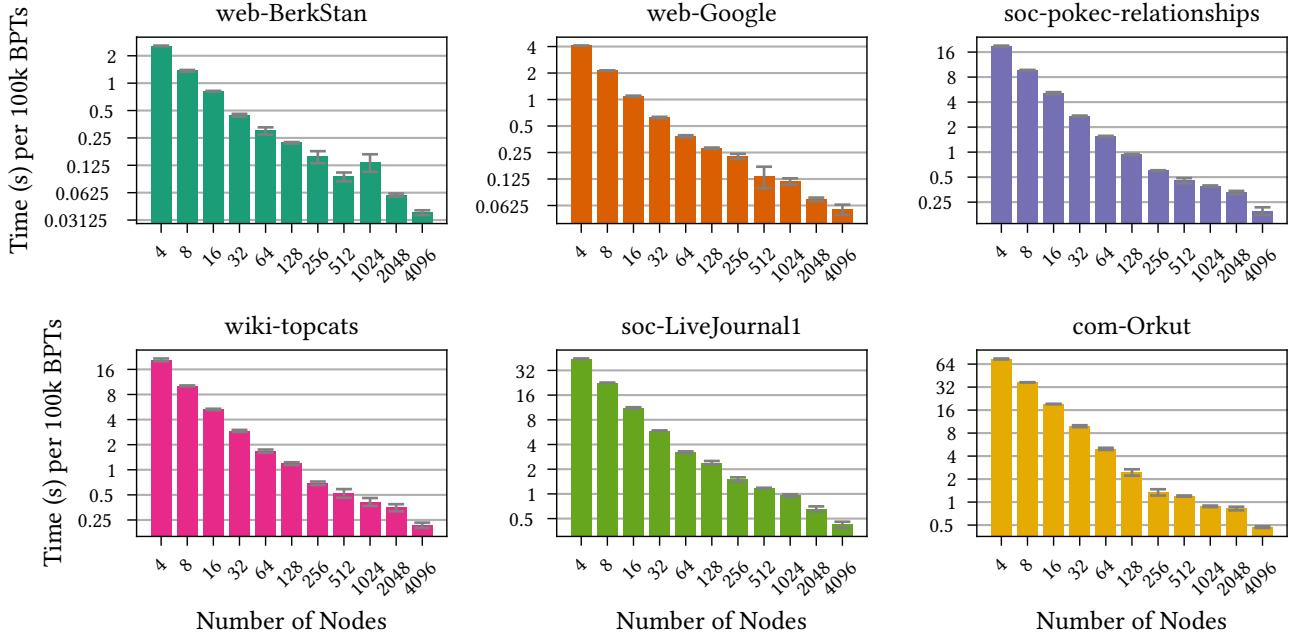}
    \caption{Multi node scaling up to 4K nodes. All times are averaged over three runs (std. dev. shown in gray).}
    \vspace{-7pt}
    \label{fig:multiscaling}
\end{figure*}


Applying the principle of fused BPTs to Ripples provides two key improvements:
\begin{enumerate*}
    \item \textsl{BPT Concurrency:} While fusing BPTs in gIM doubles the number of BPTs on the GPU at any given time, applying the approach to Ripples increases the number of BPTs on the GPU from 1 to the number of colors. For example, in the 1024 color case, this can lead to a 1024-fold increase in BPT concurrency on the GPU.
    \item \textsl{Edge Sharing:} Ripples benefits from edge sharing like gIM, as fusing BPTs provides opportunities to reduce the amount of work; however, too many colors might reduce the GPU utilization per color, and increase the overheads for processing empty colors (see \cref{sec:syntheticanalysis}).   
\end{enumerate*}

Fig.~\ref{fig:probscaling} shows how Ripples  scales with varying probabilities and increasing color size (i.e., BPTs). The speedup significantly increases with color size over the unfused (1 color) approach. This is due to increased concurrency. We also observe that traversals with lower probabilities benefit more from fusing with respect to traversals with higher probabilities for the majority of input graphs, even when considering probability of 1. Traversals with lower edge-probabilities reduce vertex activations, thus leading to smaller frontiers of the BPTs leading to lower GPU utilization. 
To better understand this behavior, we profiled the sizes of the hierarchical frontier queues.
Fig.~\ref{fig:probutilization} shows that with low probabilities and low number of colors, there are not enough wavefronts to keep an entire GCD fully utilized. A wavefront is the equivalent of a CUDA warp, representing the minimal scheduling unit in terms of parallel threads simultaneously executed by a SIMD unit (single instruction, multiple threads) of the GPU (either a GCU for AMD or a SM for NVIDIA). An AMD wavefront corresponds to 64 threads, a CUDA warp has 32 threads. This behavior also explains why, with lower probabilities, increasing the number of colors provides a larger improvement in performance: beside benefiting from the increases in concurrency of  BPTs and from edge sharing, our approach also increases the utilization of the GPU. 

\subsubsection{Scaling}
\label{sec:ripplesscaling}


 We evaluate the scaling of  Ripples with fused BPT  on multiple GPUs and multiple nodes. 

\begin{figure}[!hbt]
    \centering
    \input{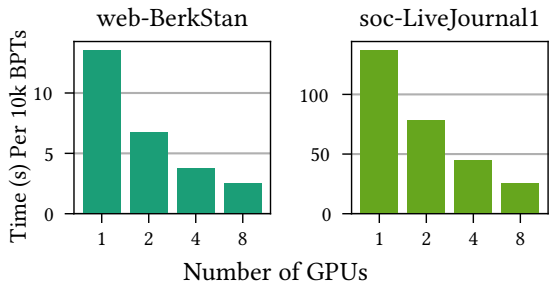}
    \vspace{-7pt}
    \caption{Single node scaling with 32 colors.}
    \label{fig:gpuscaling}
\end{figure}
\vspace{1mm}
\noindent\textbf{Single-Node Multi-GPU:}
First, we evaluate strong scaling on single node at 32 colors when progressively increasing the number of GPUs from 1 to 8. 
Fig.~\ref{fig:gpuscaling} shows that the performance of Ripples scales almost linearly when employing fused BPTs. We only report the results of two graphs, but all our bechmarks follow the same trends. Outside of the greater potential for workload imbalances due to increased workload granularity, fusing does not impact the scaling behavior of the implementation.

\vspace{1mm}
\noindent\textbf{Heterogeneous Workload Balancing:}
We then evaluate the strong scaling on multiple nodes. When increasing the number of nodes, the number of BPTs that each node processes accordingly reduces. This allowed us to identify a workload imbalance between CPU and GPU workers on the same node. CPU workers are slower than GPU ones, thus starving the GPU workers if not enough work units are available. After implementing the load balancing mechanisms described in \S\ref{sec:challenges}, we achieved a better balance between CPU and GPU workers, allowing CPUs to effectively help in the BPT generation and thus providing a speed up with respect to a GPU-only setup.


\vspace{1mm}
\noindent\textbf{Multi-Node Scaling:}
Finally, we evaluate Ripples with the fused BPT approach when increasing the number of nodes from 4 to 4096 nodes. At 4096 nodes, Ripples uses a total of 196,608 CPU cores and 32,768 GPUs for RR set generation. Fig.~\ref{fig:multiscaling} shows that our approach keeps scaling as we increase the number of nodes, reaching at 4096 nodes a speedup of $128\times$ over the 4 node version. With weak scaling, the speedup would be even higher as resources would be more utilized at higher node counts.

\vspace{-0.15in}
\section{Related Work}
\label{sec:related-work}

\noindent \textbf{Breadth-First Search (BFS)} is considered as one of the core primitives of graph algorithms as well as a prototypical irregular kernel used in benchmarking~\cite{Graph500}, BFS has received extensive attention in literature. BFS on GPU architectures has also been explored extensively \cite{Shi2018}, with notable works such as: \citet{Merrill2012} who employed hierarchical queues for fine-grained task management to scale traversals; efficient thread scheduling, degree-based sorting and direction-switching in Enterprise \cite{Liu2015}; a data-centric abstraction using bulk synchronous model to enable programming productivity in Gunrock \cite{Wang2016}; and a collection of techniques to address the data-specific challenges that are dynamic in nature in XBFS \cite{Gaihre2019}. 
Multi-threaded~\cite{Bader2006, Pearce2010},
distributed~\cite{Yoo2005}, and algebraic approaches for BFS~\cite{Yang2022} have also been studied well.


\vspace{1mm}
\noindent \textbf{Fused Breadth-First Search}
have been part of two prior implementations. MS-BFS, introduced by \citet{Then2014}, is a single-threaded CPU-only method for fusing BFSs. 
MS-BFS employs a bitmask coloring scheme, and introduces techniques to improve performance such as aggregated neighbor processing, direction-optimized travel, neighbor prefetching, and degree-bared vertex reordering. \citet{Liu2016} proposed iBFS, an extension of Enterprise \cite{Liu2015}, to perform parallel concurrent BFSs on GPUs. iBFS also utilizes a bitmask-based data structure and introduces GroupBy rules for improved sharing along with early termination for their bottom-up approach. In contrast to our work, both implementations handle deterministic traversals, allowing them to take advantage of the guaranteed traversal of all edges in the frontier. This allows for helpful heuristics such as direction-switching and immediate joining of destination vertices in the frontier. In contrast, our BPT implementation keeps edges to the same destination separate during traversal, as each edge needs to be evaluated randomly to maintain correctness.

\section{Conclusion and Future Work}
\label{sec:conclusion}
In this work, we have proposed a fused BPT algorithm, where we share frontiers between separate BPTs. We implemented our algorithm on two  frameworks, gIM and Ripples, that use two different approaches for generating BPTs. Through our experiments, we show the benefits of fusing over their unfused counterparts. We also identify source binning, vertex reordering, and workload balancing as key heuristics for improving the performance of fused BPT on single-accelerator, heterogeneous, and distributed systems. Additionally, we show strong scaling results of both single-node multi-GPU and multi-node heterogeneous systems.

Future research directions include (but are not limited to): 
\begin{enumerate*}[label=\alph*)]
    \item exploration of adaptive techniques for improved performance in probabilistic traversals---e.g., directional switching \cite{beamer2012}, hybrid scan vs. queue frontier management methods \cite{Li2013, nasre2013}, and dynamic color sizes;
    \item potential ways to pause and resume workloads so that finished BPTs can offload their RR sets early and new start nodes can be injected mid-traversal; and 
    \item exploiting any higher order structural information of a graph into fused BPTs. 
\end{enumerate*}


\bibliographystyle{ACM-Reference-Format}
\bibliography{references,halappanavar}

\end{document}